\def\c#1{[#1]}
\def\l#1{\eqno(#1)}
\def\r#1{(#1)}
\def\i#1{\par[#1].}
\def\newpage{\vfill\eject}
\def\frac#1#2{{{#1} \over {#2}}}
\def\qps#1#2{
\mathrel{\mathop{{#1}}\limits^2},
\mathrel{\mathop{{#2}}\limits^1} 
}
\def\qp{\qps{\bf x}{-i\frac{\partial}{\partial {\bf x}}} }
\pageno=0
\tolerance=250
\magnification=1200

\centerline{}

\centerline{}

\centerline
{INITIAL CONDITIONS}

\centerline{FOR SEMICLASSICAL FIELD THEORY}

hep-th/9709151

\centerline{}

\centerline{}

\centerline
{V.P.Maslov and O.Yu.Shvedov \footnote*
{e-mail: olshv@ms2.inr.ac.ru}}

\centerline{\it Chair of Quantum Statistics and Field Theory,}

\centerline{{\it Dept. of Physics, Moscow State University,}}

\centerline{{\it 119899, Moscow, Russia}}

\centerline{\it and Bristol University, Bristol, United Kingdom}

\centerline{}

\centerline{\bf Abstract}

Semiclassical approximation based on extracting  a  c-number
classical component from quantum field is widely used in the
quantum field theory.  Semiclassical states  are  considered
then  as Gaussian wave packets in the functional Schr\"odinger
representation and as Gaussian vectors in the Fock representation.  
We consider the problem of divergences and renormalization 
in the semiclassical field theory in the Hamiltonian formulation. 
Although divergences in quantum field theory are usually associated 
with loop Feynman graphs, divergences
in the Hamiltonian approach may arise even at the tree level.
For example, formally calculated probability of pair creation  
in  the leading order of the semiclassical expansion may
be divergent.This observation was interpretted as  an  argument  
for considering non-unitary evolution transformations,
as well as non-equivalent representations of canonical  commutation  
relations at different time moments.  However,  we
show that this difficulty can be overcomed without  the  assumption 
about non-unitary evolution.  We consider first the
Schrodinger equation for the regularized field  theory  with
ultraviolet  and  infrared cutoffs.  We study the problem of
making a limit to the local theory.  To consider such a  limit, 
one should impose not only the requirement on the counterterms 
entering to the quantum Hamiltonian  but  also  the
requirement on the initial state in the theory with cutoffs.
We find such a requirement in the leading order of  the  
semiclassical  expansion  and  show that it is invariant under
time evolution. This requirement is also presented as a condition 
on the quadratic form entering to the Gaussian state.


\newpage

\centerline{{\bf 1. Introduction}}

Semiclassical approximation is of widely use in quantum field theory.
Examples of applying this approximation are:

(i) qunatization in the vicinity of soliton solutions to
the classical field theory \c{1,2,3};

(ii) investigation of processes in strong external
electromagnetic and gravitational fields \c{4,5,6,7};

(iii) Gaussian approximation of refs. \c{8,9,10}.

To construct semiclassical theory, one usually starts from
considering the field model with action depending on fields
$\varphi$ and small paremeter $g$ as follows,
$$
\frac{1}{g}S[\varphi \sqrt{g}].
\l{1}
$$
Then one extracts classical part of the field which is large,
of order $1/\sqrt{g}$, according to the following formula,
$$
\varphi=\Phi_c/\sqrt{g} + \phi
\l{2}
$$
Substituting eq.\r{2} to eq.\r{1}, one finds that the leading
order of the action \r{1} is quadratic,
$$
\frac{1}{2} \phi
\frac{\delta^2 S}{\delta\Phi_c \delta \Phi_c}\phi + const,
\l{3}
$$
since the term which is linear in
$\phi$ vanishes because of the classical equations of motion.

  Since the field theory \r{3} is the theory with quadratic
Lagrangian, it is exactly solvable. Quantization of the
action \r{3} leads to the Schr\"odinger equation with quadratic
Hamiltonian depending of the creation and annihilation
operators    $a^{\pm}({\bf k})$ as follows,
$$
\hat{H}_2= 
[\frac{1}{2}a^+ \hat{A}^t a^+ +
a^+ \hat{B}^t a^-
+\frac{1}{2}a^- \hat{A}^{t*} a^-] +\gamma^t,
\l{4}
$$
where notations like $a^{\pm}\hat{O}a^{\pm}$ are used for
bilinear forms in creation and annihilation operators,
$\int d{\bf k} d{\bf p} a^{\pm}({\bf k}) O({\bf k},{\bf p})
a^{\pm}({\bf p})$, $\bf k$ and $\bf p$ are momenta of particles,
the operator with the kernel $O({\bf k},{\bf p})$ is denoted by
$\hat{O}$, $\gamma^t$ is a number.

Note that the coefficients $A^t,B^t$ are time-dependent
for the case of non-stationary classical field solutions,

To construct exact solutions to eq.\r{3}, one can use the
following Gaussian ansatz for the vector of the Fock space,
$$
 c^t \exp(\frac{1}{2}  a^+ \hat{M}^t
a^+) \Phi^{(0)},
\l{5}
$$
Eq. \r{5} obey the quadratic Schr\"odinger equation if
$$
i\dot{\hat{M}}^t=\hat{A}^t + \hat{B}^t \hat{M}^t +
\hat{M}^t (\hat{B}^t)^T + \hat{M}^t \hat{A}^{t*} \hat{M}^t.
\l{6}
$$
However, in this approach the problem of divergences and
renormalization in quantum field theory are not taken into
account. This leads to some difficulties in the semiclassical
field theory which have been studied in refs. \c{4,11}.
Namely, expression \r{5} really determines the vector of the
Fock space if and only if \c{12} the kernel of the operator
$\hat{M}^t$, $M^t({\bf k},{\bf p})$ is square integrable,
$$
\int d{\bf k} d{\bf p} |M^t({\bf k},{\bf p})|^2 < \infty,
\l{7}
$$
because the two-particle component of the vector \r{5}
that corresponds to creation of the single pair of particles
with momenta $\bf k$ and $\bf p$ is $M^t({\bf k},{\bf p})$,
so that expression \r{7} is the full probability of
pair creation. This probability should be finite. However,
evolution of the quantity \r{7} is prescripted by eq.\r{6}.
It was found in refs.\c{11,4} that for some field models
like quantum gravity or QED the quantity \r{7} may be
divergent even if it converged at the initial time moment.

This difficulty of semiclassical theory lead to the assumption
that time evolution in quantum field theory is non-unitary,
so that one should consider non-equivalent representations of
the canonical commutation relations \c{11,13}.

We will show in this paper that this difficulty can be overcomed
in the leading order of the semiclassical expansion
without the assumption about non-unitary evolution.

We will consider the renormalization procedure in the Hamoiltonian
formulation. It will be shown that state vectors of the linearized
quantum theory \r{3} are not in one-to-one correspondence
with elements of the renormalized state space. This means that not
the condition \r{7} but another condition should be imposed
on the function $M^t({\bf k},{\bf p})$. Namely, the function
$M^t({\bf k},{\bf p})$ is the probability amplitude to emit
{\it bare} particles. However, {\it physical} particles are
not identical to bare particles, because one should consider also
the processes of self-interaction (see, for example, \c{14});
there is ''dressing transformation'' that transforms bare
particles to physical particles. We will analyse this
transformation and find the more complicated condition
on the quadratic form entering to the Gaussian state \r{5}
which will provide the convergence of all probabilities
computated for physical particles.

This paper is organized as follows. Section 2 deals with semiclassical
approximation for the field theory with ultraviolet and
infrared cutoffs. Since the divergences do not arise in such a
regularized theory, one can investigate the semiclassical
asymptotics rigorously. We will also associate this semiclassical
approach  with the functional - Schr\"odinger - equation variational
approach suggested in \c{8,9,10}. It will happen that
these approaches are equivalent at small values of $g$.
In section 3 we will consider the problem of making limit to the
local field theory. We will study conditions on the regularized
Hamiltonian and on the possible states which are necessary to
make the renormalized theory finite in the leading order of
the semiclassicla expansion.
We will construct ''dressing transformation'' by using the
Bogoliubov method \c{15,16} of variable intensity of the
interaction. We will find the correct condition on $M^t$
instead of \r{7}. Section 4 deals with the direct
analysis of eq.\r{6}. We will find that the obtained
condition on $M^t$ is invariant under time evolution,
so that it is not necessary to consider non-unitary
evolution or non-equivalent representations of the canonical
commutation relations at different time moments.
Section 5 contains conclusion remarks and discussions of the
obtained condition on $M^t$.

\centerline{{\bf 2. Semiclassical approximation for the
regularized field theory}}

This section deals with constructing the semiclassical asymptotics 
in quantum field theory. For the simplicity, we consider in
this paper the case of the single scalar field  only, because
other cases can be considered analogously. We will study the 
action of the form
$$
S[\varphi,g]= \int dx [\frac{1}{2} \partial_{\mu} \varphi
\partial^{\mu} \varphi -
\frac{m^2}{2} \varphi^2  -
\frac{1}{g} V_{int}(\sqrt{g}\varphi)],
\l{8}
$$
where $x=(x^0,...,x^d)$, $V_{int}(\Phi)=O({\Phi}^3)$,
$\partial_{\mu}= \partial/\partial x^{\mu}$.

It is well-known that quantization of the field theory with the action
\r{8} lead to ultraviolet and volume divergences. Therefore,
one usually studies the action with ultraviolet and infrared
cutoffs instead of \r{8}.  
The regularized action is
$$
S^0_{\Lambda,L}[\varphi,g]=
\int dx [\frac{1}{2} \partial_{\mu} \varphi
\partial^{\mu} \varphi
- \frac{m^2}{2}\varphi^2]
- \int dx \chi({\bf x}/L)
\frac{1}{g} V_{int}(\sqrt{g}\varphi^{\Lambda}(x^0,{\bf x})],
\l{9}
$$
where
$$
\varphi^{\Lambda}(t,{\bf x}) = \int d{\bf y}
\rho(\Lambda({\bf x} - {\bf y}))
\varphi^{\Lambda}(t,{\bf y}),
$$
$\chi$ and $\rho$ are smooth and rapidly damping functions, $\chi(0)=1$,
$\int d{\bf z} \rho({\bf z})=1$.
If $\Lambda,L \to \infty$ then action \r{9}
takes the form \r{8}.

However, the corresponding quantum field theory is not well-defined as
$\Lambda,L \to \infty$.
To make this limit, $\Lambda,L \to \infty$, possible, one should add
to \r{9} the counterterms:
$$
S^0_{\Lambda,L}+ S^{ct}_{\Lambda,L},
\l{10}
$$
where the added term $S^{ct}_{\Lambda,L}(\varphi,g)$ depends on the 
small parameter $g$ like this:
$$S^{ct}(\varphi,g)=\sum_{n\ge 1} g^{n-1}
S^n_{\Lambda,L}(\varphi\sqrt{g})$$.
The purpose of the 
functional $S^{n}_{\Lambda,L}$
is to remove the $n$-loop divergences.

Let us consider quantization of the field theory with the action
\r{10} and constructing semiclassical asymptotics in different
representations.

Classical Hamiltonian corresponding to the action \r{10},
depends on the fields $\varphi({\bf x})$ and canonically
conjugated momenta $\pi({\bf x})$. It can be presented
as a sum
$$
H_{\Lambda,L}=H^0_{\Lambda,L} + H^{ct}_{\Lambda,L}
$$
of the functional
$$
H_{\Lambda,L}^0=
\frac{1}{g}H_{\Lambda,L}^{(0)}[\pi\sqrt{g},\varphi\sqrt{g}]=
\frac{1}{g}H_{0}[\pi\sqrt{g},\varphi\sqrt{g}]+
\frac{1}{g}H_{\Lambda,L}^{int}[\pi\sqrt{g},\varphi\sqrt{g}],
\l{11}
$$
where
$$
H_0[\Pi(\cdot),\Phi(\cdot)]= \int
d{\bf x} [\frac{1}{2}\Pi^2({\bf x}) + \frac{1}{2}
(\nabla \Phi)^2({\bf x}) + \frac{m^2}{2}\Phi^2({\bf x})],
$$
$$
H_{\Lambda,L}^{int}[\Pi(\cdot),\Phi(\cdot)]=
\int d{\bf x} \chi({\bf x}/L) V_{int}(\Phi^{\Lambda}({\bf x})),
$$
and the counterterm Hamiltonian depending on the small parameter $g$ as
$$
 H^{ct}_{\Lambda,L}= \sum_{n\ge 0}
g^n H_{\Lambda, L}^{(n+1)} [\pi\sqrt{g},\varphi\sqrt{g}].
$$
We will define the form of $H_{\Lambda, L}^{(n+1)}$
later.

The procedure of canonical quantization is well-known.
One should substitute fields $\varphi$ and momenta
 $\pi$ by the operators $\hat{\varphi}$ and $\hat{\pi}$,
which act in the Hilbert state space $\cal H$,
 and obey the canonical commutation relations,
$$
[\hat{\varphi}({\bf x}),\hat{\varphi}({\bf y})]=0,
[\hat{\pi}({\bf x}),\hat{\pi}({\bf y})]=0,
[\hat{\varphi}({\bf x}),\hat{\pi}({\bf y})]=i\delta({\bf x}-{\bf y}),
\l{12}
$$
Evolution of the state is prescribed by the Schr\"odinger 
equation
$$
i\frac{d}{dt} \Psi^t_{\Lambda,L,g}=
\sum_{n\ge 0} g^{n-1}  H^{(n)}_{\Lambda,L}
[\sqrt{g}\hat{\pi}(\cdot),\sqrt{g}\hat{\varphi}(\cdot)]
\Psi^t_{\Lambda,L,g}, \Psi^t_{\Lambda,L,g} \in {\cal H}
\l{13}
$$
One usually chooses the Fock space as the space $\cal H$. 
Operators $\hat{\pi}$ and $\hat{\varphi}$
are presented through the creation and annihilation operators as follows
$$
\matrix{
\hat{\pi}({\bf x})=\frac{i}{(2\pi)^{d/2}}
\int d{\bf k} \sqrt{\frac{\omega_{\bf k}}{2}}
[a^+({\bf k})e^{-i{\bf kx}} - a^-({\bf k}) e^{i{\bf kx}}],
\cr
\hat{\varphi}({\bf x})
=\frac{1}{(2\pi)^{d/2}}
\int d{\bf k} \sqrt{\frac{1}{2\omega_{\bf k}}}
[a^+({\bf k})e^{-i{\bf kx}} + a^-({\bf k}) e^{i{\bf kx}}],
}
\l{14}
$$
where $\omega_{\bf k}= \sqrt{{\bf k}^2+m^2}$.

Since the operators $\hat{\pi}$ and $\hat{\varphi}$ are non-commutative
according to eq.\r{12}, there is a problem to order the operators:
it is not clear how to construct operators
$H_{\Lambda, L}^{(n)} [\hat{\pi}(\cdot)\sqrt{g},
\hat{\varphi}(\cdot)\sqrt{g}]$,
using the functionals
$H_{\Lambda, L}^{(n)} [{\pi}(\cdot)\sqrt{g},{\varphi}(\cdot)\sqrt{g}]$
In quantum field theory one ususally uses the Wick ordering: 
the operators of fields and momenta are expressed through the
creation and annihilation operators according to eq.\r{14},
then the creation and annihilation operators are ordered in
such a way that creation operators act later than annihilation
operators.

Let us consider construction of semiclassical solutions to eq.\r{3}.

In quantum mechanics, the well-known semiclassical approach
is the WKB-approach, which enables one to construct semiclassical
solutions to the Schr\"odinger equation 
$$
i\hbar\frac{\partial \psi(x,t)}{\partial t}=
H(x, -i\hbar\frac{\partial}{\partial x}) \psi(x,t)
\l{15}
$$
of the rapidly oscillating form
$$
\psi(x,t) \sim \phi(x,t) \exp(\frac{i}{\hbar}S(x,t)),
\l{16}
$$
where action $S$ is a real function. One can easily show that 
the wave function \r{16} really obeys approximately the 
Schr\"odinger equation \r{15} as $\hbar\to 0$.
However, the form \r{16} is not the only possible 
ansatz of the wave function, obeying approximately eq.\r{15} in the 
semiclassical approximation, $\hbar\to 0$. One can construct
other, non-WKB, semiclassical solution to \r{15}.

Another example of semiclassical wave function is the wave-packet
approximate solution to eq.\r{15}:
$$
\psi(x,t) \sim e^{\frac{i}{\hbar}S(t)} e^{\frac{i}{\hbar}P(t)(x-Q(t))}
f(t,\frac{x-Q(t)} {\sqrt{\hbar}} )
\l{17}
$$
corresponding to the particle moving on the classical trajectory
$(P(t),Q(t))$. 
The wave function \r{17} can be considered as
$$
U^{\hbar}_{P(t),Q(t),S(t)} \chi^{\hbar},
\l{18}
$$
where 
$$
\chi^{\hbar}(x,t)=\chi(t, x/\sqrt{\hbar}),
\l{19}
$$
while $U^{\hbar}_{P,Q,S}$ is an unitary operator of the form
$$
U^{\hbar}_{P,Q,S}=
 e^{\frac{i}{\hbar}S}
 e^{\frac{i}{\hbar}Px}
 e^{-\frac{i}{\hbar}Q(-ih\frac{\partial}{\partial x})}.
\l{20}
$$
One can show by using the complex-WKB (or complex germ)
technique of ref.\c{17} that eq.\r{17} really satisfies
eq.\r{15} as $\hbar\to 0$. Wave function \r{17} have been also 
considered as a test function in the variational method of 
refs.\c{18,8,9}.

It is the semiclassical wave function \r{17} rather than the
WKB-function \r{15} that generalizes to the secondary-quantized
systems \c{19,20}. Quantum-field analog of the complex-WKB
method is the soliton quantization.

Let us construct quantum-field-theory analog of the asymptotics
\r{17}.

We consider the more general case of the Hamiltonian
 $H_{\Lambda,L}^{(0)}$ of the form
$$
\frac{1}{g}H_{\Lambda,L}^{(0)}[\pi\sqrt{g},\varphi\sqrt{g}]=
\frac{1}{g}H_{0}[\pi\sqrt{g},\varphi\sqrt{g}]+
\xi(t) \frac{1}{g}H_{\Lambda,L}^{int}[\pi\sqrt{g},\varphi\sqrt{g}],
\l{21}
$$
where $\xi(t)$ is a smooth function of $t$.
If $\xi(t) = 1$ then the operator \r{21} takes the form
\r{11}.

Note that the small parameter $g$ is an analog of the Planck constant
$\hbar$, operators $\hat{\varphi}\sqrt{g}$ and $\hat{\pi}\sqrt{g}$
are analogs of the quantum mechanical coordinate and momentum
operators, $x$ and $-i\hbar \partial/\partial x$.
The $g$-independent state vector is an analog of the
wave function \r{19}, while the
 unitary transformation
$$
U^g_{\Phi,\Pi,S}= e^{\frac{i}{g}S}e^{\frac{i}{\sqrt{g}}
\int d{\bf x} [\Pi({\bf x})\hat{\varphi}({\bf x}) -
\Phi({\bf x})\hat{\pi}({\bf x})] },
$$
where $S$ is a  real number, $\Phi(\cdot)$, $\Pi(\cdot)$ are
real functions, is the analog of eq.\r{20}.
 It follows from the canonical commutation
relations \r{12} that
$$
\matrix{
(U^g_{\Phi,\Pi,S})^{-1} \hat{\varphi}({\bf x}) U^g_{\Phi,\Pi,S}
= \hat{\varphi}({\bf x}) + \Phi({\bf x})/\sqrt{g},
\cr
(U^g_{\Phi,\Pi,S})^{-1} \hat{\pi}({\bf x}) U^g_{\Phi,\Pi,S}
= \hat{\pi}({\bf x}) + \Pi({\bf x})/\sqrt{g},
\cr
(U^g_{\Phi,\Pi,S})^{-1} \frac{d}{dt} U^g_{\Phi,\Pi,S}
=D_t
}
\l{22}
$$
where
$$
iD_t=-\frac{1}{g}\left(\dot{S}^t+
\frac{1}{2}
\int d{\bf x} (\dot{\Pi}^t({\bf x}) \Phi^t({\bf x})-
\dot{\Phi}^t({\bf x}) \Pi^t({\bf x}) )\right)
$$
$$
+\frac{1}{\sqrt{g}} \int d{\bf x}
(\dot{\Phi}^t({\bf x}) \hat{\pi}({\bf x}) -
\dot{\Pi}^t({\bf x}) \hat{\varphi}({\bf x}) ) + id/dt.
$$
Consider the following element of the space  $\cal H$,  
$$
\Psi^t=U^g_{\Phi^t,\Pi^t,S^t} Y^t,
\l{23}
$$
where state vector $Y^t$ is regular as $g\to 0$. Subsituting 
state vector \r{23} to eq.\r{13}, making use of 
commutation rules \r{22}, one obtains the following  
equation on $Y^t$:
$$
iD_t Y^t = \sum_{n\ge 0} g^{n-1}
H^{(n)} [\Pi(\cdot)+\sqrt{g}\hat{\pi}(\cdot),
\Phi(\cdot)+\sqrt{g}\hat{\varphi}(\cdot)]Y^t,
\l{24}
$$
where indices $\Lambda,L$ are omitted.

Eq.\r{24} contains singular as $g\to 0$
terms of orders   $O(1/g)$ and $O(1/\sqrt{g})$. In order to
obtain a regular equation for $Y^t$, one should make these
terms equal to 0. One finds the following relation 
$$
\dot{S}^t=
\frac{1}{2}
\int d{\bf x} (-\dot{\Pi}^t({\bf x}) \Phi^t({\bf x})+
\dot{\Phi}^t({\bf x}) \Pi^t({\bf x}) ))
- H^{(0)}[\Pi(\cdot),\Phi(\cdot)].
\l{25}
$$
and the Hamiltonian system
$$
\dot{\Phi}^t({\bf x})= \frac{\delta H^{(0)}}{\delta \Pi({\bf x})},
\dot{\Pi}^t({\bf x})= -\frac{\delta H^{(0)}}{\delta \Phi({\bf x})}.
\l{26}
$$
Eqs.\r{25} and \r{26} imply the following 
equation on $Y^t$ as
$g\to 0$,
$$
i\frac{d}{dt}Y^t=H_2[\hat{\pi}(\cdot),\hat{\varphi}(\cdot)] Y^t,
\l{27}
$$
where quantum operator 
$\hat{H}_2=H_2[\hat{\pi}(\cdot),\hat{\varphi}(\cdot)]$
corresponds to the classical functional
$$
H_2[{\pi}(\cdot),\varphi(\cdot)]=
\frac{1}{2}\pi \frac{\delta^2 H^{(0)}}{\delta \Pi \delta \Pi}\pi+
\pi \frac{\delta^2 H^{(0)}}{\delta \Pi \delta \Phi}\varphi+
\frac{1}{2}\varphi \frac{\delta^2 H^{(0)}}{\delta \Phi \delta \Phi}
\varphi+H^{(1)}.
\l{28}
$$
the Wick ordering is assumed for
$H_2[\hat{\pi}(\cdot),\hat{\varphi}(\cdot)]$.
In eq.\r{28} the arguments   $\Phi({\cdot}),\Pi({\cdot})$
of the functionals $H^{(0)}$  and  $H^{(1)}$ are omitted, 
while integrals like
 $\int \pi({\bf x}) \frac{\delta^2 H^{(0)}}
{\delta \Pi({\bf x}) \delta \Pi({\bf y}) }\pi({\bf y})
d {\bf x}d {\bf y}$ are denoted as
$\pi \frac{\delta^2 H^{(0)}}{\delta \Pi \delta \Pi}\pi$.

Note that an alternative way to derive the  Schr\"odinger
equation \r{27} for the linearized field theory is to extract
the classical part of the field according to eq.\r{2}
and to quantize the Lagrangian \r{3} canonically.

Let us consider exact solutions to eq.\r{27}
being of the type \r{4}, where
$$
\hat{A}^t({\bf k},{\bf p})=
\int \frac{d{\bf x} d{\bf y}}{(2\pi)^d}
e^{-i{\bf kx}} \frac{1}{4}
\left(\frac{1}{\sqrt{\omega_{\bf k}}}
\frac{\delta^2 H^{(0)}}{\delta \Phi({\bf x}) \delta \Phi({\bf y})}
\frac{1}{\sqrt{\omega_{\bf k}}}
\right.
$$
$$
\left.
-
{\sqrt{\omega_{\bf k}}}
\frac{\delta^2 H^{(0)}}{\delta \Pi({\bf x}) \delta \Pi({\bf y})}
{\sqrt{\omega_{\bf k}}}
\right)
e^{-i{\bf py}},
$$
$$
\hat{B}^t({\bf k},{\bf p})=
\int \frac{d{\bf x} d{\bf y}}{(2\pi)^d}
e^{-i{\bf kx}} \frac{1}{4}
\left(\frac{1}{\sqrt{\omega_{\bf k}}}
\frac{\delta^2 H^{(0)}}{\delta \Phi({\bf x}) \delta \Phi({\bf y})}
\frac{1}{\sqrt{\omega_{\bf k}}}
\right.
$$
$$
\left.
+
{\sqrt{\omega_{\bf k}}}
\frac{\delta^2 H^{(0)}}{\delta \Pi({\bf x}) \delta \Pi({\bf y})}
{\sqrt{\omega_{\bf k}}}
\right)
e^{i{\bf py}},
\l{29}
$$
while $\gamma^t=H^{(1)}$. We have taken into account
that
$\frac{\delta^2 H^{(0)}}{\delta \Pi \delta \Phi}=0$.

Eq.\r{4} hash solutions of the form \r{5}, where
\r{6} is satisfied, while
$$
i\dot{c}^t= (Tr \hat{A}^t\hat{M}^t + H^{(1)}) c^t.
\l{30}
$$
To construct other exact solutions to eq.\r{4} which obey
initial conditions of the type
$$P(a^+)
\exp(\frac{1}{2}  a^+ \hat{M}^0
a^+) \Phi^{(0)},$$ where
$P(a^+)$ is a polynomial in creation operators, one can
introduce the so-called complex germ (or complex-WKB)
creation operators of ref.\c{17}:
$$
\Lambda[p^t,q^t]=\int d{\bf x} [p^t({\bf x})\hat{\varphi}({\bf x})-
q^t({\bf x}) \hat{\pi}({\bf x})]
\l{31}
$$
Because of the canonical commutation relations
$$
[i\frac{d}{dt}  - \hat{H}_2 , \Lambda[p^t,q^t]]
= i \Lambda[\dot{p}^t+\frac{\delta H_2}{\delta q},
\dot{q}^t-\frac{\delta H_2}{\delta p}]
\l{32}
$$
operator \r{31} transforms a solution to eq.\r{27} into a solution
if
$$
\dot{p}^t({\vec x}) = - \frac
{\delta H_2(p^t(\cdot),q^t(\cdot))}{\delta q({\bf x})},
\dot{q}^t({\vec x}) =  \frac
{\delta H_2(p^t(\cdot),q^t(\cdot))}{\delta p({\bf x})}.
\l{33}
$$

Let us simplify eqs.\r{6} and \r{30}.
Consider the operator $\tilde{M}^t$ with the kernel
being the Fourier transformation of the kernel 
$M^t({\bf k},{\bf p})$ of 
the operator $\hat{M}^t$,
$$
(\tilde{M}^t f)({\bf x})=
\int \frac{d{\bf k} d{\bf p} d{\bf y}}{(2\pi)^d}
e^{i({\bf kx} + {\bf py})} M^t({\bf k},{\bf p})
 f({\bf y}),
\l{34}
$$
Making use of this substitution \r{34}, 
one finds that
$$
i\dot{\tilde{M}}^t=
\frac{1}{2}(1+\tilde{M}^t)
\frac{1}{\sqrt{\hat{\omega}}}
\frac{\delta^2 H^{(0)}}
{\delta \Phi \delta \Phi }
\frac{1}{\sqrt{\hat{\omega}}}
(1+\tilde{M}^t)
-
\frac{1}{2}(1-\tilde{M}^t)
{\sqrt{\hat{\omega}}}
\frac{\delta^2 H^{(0)}}
{\delta \Pi \delta \Pi }
{\sqrt{\hat{\omega}}}
(1-\tilde{M}^t)
\l{35}
$$
where $\hat{\omega}$ is the operator
$\hat{\omega}=\sqrt{-\Delta+m^2}$.
while $\frac{\delta^2 H^{(0)}}{\delta \Phi \delta \Phi }$
and  $\frac{\delta^2 H^{(0)}}{\delta \Pi \delta \Pi }$
are operators with the kernels
$\frac{\delta^2 H^{(0)}}
{\delta \Phi({\bf x}) \delta \Phi({\bf y}) }$
and
$\frac{\delta^2 H^{(0)}}
{\delta \Pi({\bf x}) \delta \Pi({\bf y}) }$
correspondingly.
Eqs.\r{30} can be written as
$$
(ln c^t)\dot{}= -\frac{i}{4}
[(
\frac{1}{\sqrt{\hat{\omega}}}
\frac{\delta^2 H^{(0)}}
{\delta \Phi \delta \Phi }
\frac{1}{\sqrt{\hat{\omega}}}
-
{\sqrt{\hat{\omega}}}
\frac{\delta^2 H^{(0)}}
{\delta \Pi \delta \Pi }
{\sqrt{\hat{\omega}}}
) \tilde{M}^t]
-i H^{(1)}.
\l{36}
$$

Equation for the quadratic form entering to the Gaussian
state vector, can be simplified if one considers the 
functional Schr\"odinger representation of refs.\c{8,9}
instead of the Fock representation.
In the functional representation, states are presented as
functionals 
 $\Psi[\phi(\cdot)]$ of a real function 
 $\phi({\bf x})$, operators
$\hat{\varphi}({\bf x})$ are presented as operators
of multiplication by
$\phi({\bf x})$, operators $\pi({\bf x})$ are differential operators,
 $-i\delta/\delta \phi({\bf x})$.

Let us transform the Gaussian vector \r{5} of the Fock space into the
Schr\"odinger representation. Notice that the vector
\r{5}
is determined uniquely from the relation
$$
(a^-({\bf k})- (\hat{M}^t a^+)({\bf k})) Y =0.
\l{37}
$$
which can be easily transformed into the
Schr\"odinger representation, because the creation and annihilation 
operators can be expressed through the field and momenta operators.
Namely, the operators
$$
\tilde{a}^{\pm}({\bf x})=
\int \frac{d{\bf k}}{(2\pi)^{d/2}}
a^{\pm}({\bf k}) e^{\mp i {\bf kx}},
$$
are expressed as follows,
$$
\tilde{a}^{\pm}= \sqrt{\frac{\hat{\omega}}{2}}\hat{\varphi}
\mp i \frac{1}{\sqrt{2\hat{\omega}}}\hat{\pi}.
$$
so that eq.\r{37} can be presented as
$$
\left[(1-\tilde{M})
 \sqrt{\frac{\hat{\omega}}{2}} \phi
- (1+\tilde{M})
\frac{1}{\sqrt{2\hat{\omega}}}
\frac{\delta}{\delta \phi}\right] Y[\phi(\cdot)]=0.
$$
The solution to this equation has
the Gaussian form,
$$
Y[\phi(\cdot)]=c_q^t \exp\left[\frac{i}{2}\int
d{\bf x} \phi({\bf x}) (R\phi)({\bf x})\right]
\l{38}
$$
where
$$
R=i\sqrt{\hat{\omega}}(1-\tilde{M})(1+\tilde{M})^{-1}
\sqrt{\hat{\omega}}
\l{39}
$$
These Gaussian functionals were considered as test functionals of
the variational method of
 \c{8,9}.

The equation for the 
quadratic form takes the form
$$
\dot{R}+R^2+
\frac{\delta^2 H^{(0)}}
{\delta \Phi \delta \Phi }=0,
\l{40}
$$
where the relation
$
\frac{\delta^2 H^{(0)}}
{\delta \Pi \delta \Pi }=1
$
is taken into account. Making  use of the following 
transformation of the pre-exponential factor,
$$
c^t_q = \frac{c^t}{\sqrt{det(1+\tilde{M})}}
\l{41}
$$
one finds from eq.\r{30} that
$$
(ln c_q^t)\dot{}= -\frac{1}{2}
Tr (R-i\sqrt{-\Delta+m^2}-\frac{i}{2\sqrt{-\Delta+m^2}}
\frac{\delta^2 H^{(0)}}
{\delta \Phi \delta \Phi }) -iH^{(1)}.
\l{42}
$$
Note that eqs.\r{40},\r{42} can be found by
direct substituting of the functional  \r{38}
to eq.\r{27}. 
The relation between pre-exponential factors
in eqs. \r{5} and \r{38} can be obtained by 
expressing the 
vector \r{5} through the coherent states
according to \c{12} and using the
well-known formulas for the coherent state in the functional
representation.

Let us substract the argument of the complex factor
$c^t_q$.
Consider the real and imaginary parts of the operator
$R$, which are Hermitian. Since the imaginary part of $R$
satisfies the equation
$
(Im R)\dot{}= - Re R Im R - Im R Re R,
$
one has
$(ln det Im R)\dot{}=-2 Tr Re R$.
Thus, the quantity
$$
a^t=c_q^t/(det Im R)^{1/4}=
\frac{c^t}{(det(1+\tilde{M}))^{1/2} (det Im R)^{1/4}}
\l{43}
$$
satisfies the equaton
$$
i(ln a^t)\dot{}=
-\frac{1}{2}Tr (Im R - \sqrt{-\Delta+m^2} -
\frac{1}{2\sqrt{-\Delta+m^2}}
\frac{\delta^2 H^{(0)}}
{\delta \Phi \delta \Phi }) -H^{(1)}.
\l{44}
$$

\centerline{{\bf 3. The problem of renormalization in the 
Hamiltonian formulation}}

Let us consider the semiclassical formulas in the limit
 $\Lambda \to\infty$,
$L\to \infty$. This problem is non-trivial because the 
evolution operator is singular in this limit, contrary to the
S-matrix: formal Schr\"odinger equation in the local 
field theory is not well-defined because of the
Stueckelberg divergences \c{21}.

In the constructive field theory (see, for example,
\c{22,23,24,25}) this limit is considered as follows. 
One considers the ''renormalized'' state space
${\cal H}_{ren}$  and unitary operator
$T_{\Lambda,L}:{\cal H}_{ren} \to {\cal H}$,
depending on the parameters of the ultraviolet and infrared cutoffs.
Let $U^t_{\Lambda,L}$ be the evolution operator in the
regularized theory.
 The operator $T_{\Lambda,L}$ which is usually called as
 ''dressing transformation'' is shoosen in order to make the  
 operators
$$
W^t_{\Lambda,L}=(T_{\Lambda,L})^{-1} U^t_{\Lambda,L}
 T_{\Lambda,L}
\l{45}
$$
regular as $\Lambda,L \to \infty$. 
The operator 
$\lim_{\Lambda,L \to \infty} W^t_{\Lambda,L}$ is the
renormalized evolution operator.

This definition means the following.
If the initial condition
for eq.\r{13} is of the form
$$
T_{\Lambda,L}X,
\l{46}
$$ where
$X$ is $\Lambda,L$-independent renormalized state  
vector from ${\cal H}_{ren}$, 
then the solution to the Schr\"odinger 
equation is written in analogous way,
$$
T_{\Lambda,L}X^t+Z^{t}_{\Lambda,L},
$$
where $Z^{t}_{\Lambda,L}$ tends to zero as
$\Lambda,L \to \infty$.

Note that if the operator expressions 
$(T_{\Lambda,L})^{-1} a^{\pm}({\bf k}) T_{\Lambda,L}$
are regular as $\Lambda,L \to \infty$ then the transformation
$T_{\Lambda,L}$ allows us to construct a representation of
the canonical commutation relations. This representation may
be of the Fock or of the non-Fock type.

However, the transformations $T_{\Lambda,L}$ have been constructed
beyond perturbation theory only for some simple cases, namely,
for two-dimensional and three-dimensional models
\c{22,23,24,25}. For general quantum field models
the operator $T_{\Lambda,L}$ can be constructed only as a formal
perturbation series in $g^{1/2}$ (examples are cited
in refs.\c{14,23,26}). One can prove then that 
the coefficients of the expansion of the operator \r{45}  
are regular.

Let us define the notion of the semiclassically-regular operator \r{45}.
We will say that the operator is semiclassically regular if 
the parameters entering to the constructed 
semiclassical asymptotics of the vector
$W^t_{\Lambda,L}\Psi_{\Lambda,L}^g$ have limits as
$\Lambda,L \to \infty$. This means that for all
pairs of smooth and rapidly damping functions $\Phi^0({\bf x})$, $
\Pi^0({\bf x})$, there exists the pair of functions
$\Phi^0_{\Lambda,L}({\bf x}), \Pi^0_{\Lambda,L}({\bf x})$
and number $S^t$ that the operator
$$
V^t_{\Lambda,L,g}=
(U^g_{\Phi_{\Lambda,L}^t,\Pi_{\Lambda,L}^t,S_{\Lambda,L}^t})^{-1}
 W^t_{\Lambda,L,g}
U^g_{\Phi^0,\Pi^0,0}
\l{47}
$$
has a limit as $g\to 0$, while the operators
$$
\lim_{g\to 0}
(U^g_{\Phi_{\lambda,L}^t,\Pi_{\Lambda,L}^t,S_{\Lambda,L}^t})^{-1}
 W^t_{\Lambda,L,g}
U^g_{\Phi^0,\Pi^0,0}
\l{48}
$$
are regular as $\Lambda,L \to \infty$.
We will call the operator
$T_{\Lambda,L}$ entering to eq.\r{45} as the ''semiclassical 
dressing transformation''.

Let us construct such an operator, making use of the 
Bogoliubov method of variable intensity of the interaction
\c{15,16} (see also \c{27}).

Consider the smooth function 
$\xi(\tau)$, $\tau \le 0$ vanishing as $\tau < -T_1$
and being equal to 1 at $\tau > -T_2$ (fig.1). Consider the 
operator
$$
H_{int}^{\Lambda,L} =
\xi(\tau) \int d{\bf x} \chi({\bf x}/L)
\frac{1}{g} V_{int}(\sqrt{g}\varphi^{\Lambda}({\bf x}))+
H_{ct}(\sqrt{g}\hat{\varphi}(\cdot),\sqrt{g}\hat{\pi}(\cdot),
\tau,\xi(\cdot))
\l{49}
$$
When $\tau \in (-T_2,0)$, the functional $H_{ct}$ is 
$\tau$-independent,  while $H_{ct} =0$ as $\tau < -T_1$ 
The counterterm $H_{ct}$ is to be constructed in the next
section.

Let $H_0=\int d{\bf k} \omega_{\bf k} a^+({\bf k})
a^-({\bf k})$ be a free Hamiltonian. Consider the operator
$$
T^{\Lambda,L}=T\exp[-i\int_{-\infty}^0 d\tau e^{iH_0\tau}
H^{\Lambda,L}_{int} (\tau) e^{-iH_0\tau}]
\l{50}
$$
which transforms the initial condition of the Cauchy problem
$$
\matrix{
i\frac{dX^{\tau}}{d\tau}=e^{iH_0\tau}
H^{\Lambda,L}_{int} (\tau) e^{-iH_0\tau} X^{\tau}
\cr
X^{-\infty}=X
}
\l{51}
$$
to the solution to this problem as $\tau=0$. 
The right-hand side of 
eq.\r{51} vanish at
$\tau < -T_1$, so that the initial condition
at $\tau=-\infty$
can be substituted by the initial condition at
$\tau=-T_0<-T_1$.

Consider the semiclassical approximation for the vector
\r{46}.

Introduce the following notation.
Let   $S$ and $ln a$ be real numbers, $\Phi({\bf x})$,
$\Pi({\bf x})$, $q({\bf x})$, $p({\bf x})$ be smooth and rapidly damping 
at the infinity functions, $R$ be an operator with symmetric kernel
and positively defined imaginary part.
Consider these objects as initial conditions for the 
dynamical equations
 \r{26}, \r{25}, \r{44}, \r{33}, \r{40} at
$t=t_1$:
$$
p^{t_1}=p, q^{t_1}=q, \Phi^{t_1}=\Phi,
\Pi^{t_1}=\Pi, R^{t_1}=R, S^{t_1}=S, a^{t_1}=a.
\l{52}
$$
Consider the solutions of these equations at $t=t_2$:
$$
p^{t_2}, q^{t_2}, \Phi^{t_2},\Pi^{t_2}, R^{t_2}, S^{t_2}, a^{t_2}.
\l{53}
$$
Denote this transformation as
 ${\cal U}^{t_2,t_1}_{\xi(\cdot)}$:
$$
 (p^{t_2}, q^{t_2}, \Phi^{t_2},\Pi^{t_2}, R^{t_2}, S^{t_2}, a^{t_2})=
{\cal U}^{t_2,t_1}_{\xi(\cdot)}
(p^{t_1}, q^{t_1}, \Phi^{t_1},\Pi^{t_1}, R^{t_1}, S^{t_1}, a^{t_1})
$$
where $\xi(\cdot)$ enter to eq.\r{21}.

To construct state     $T^{\Lambda,L}X^{-\infty}$ as
$g\to 0$, consider the substitution
 $\Psi^t=e^{-iH_0\tau} X^{\tau}$
for eq.\r{51}. The Cauchy problem \r{51} takes the form
$$
\matrix{
i\frac{d\Psi^{\tau}}{d\tau}=[H_0+
H^{\Lambda,L}_{int} (\tau)]\Psi^{\tau}
\cr
\Psi^{-T_0}=e^{iH_0T_0}X^{-\infty}
}
\l{54}
$$
Let us consider the initial condition $X^{-\infty}$ 
$$
\matrix{
X^{-\infty}_g=U^g_{\Phi,\Pi,0} c \Lambda[p_1(\cdot),q_1(\cdot)]
...\Lambda[p_k(\cdot),q_k(\cdot)]
\cr
\times
\exp(\frac{1}{2}\int d{\bf k} a^+({\bf k})(Ma^+)({\bf k}))\Phi^{(0)},
}
\l{55}
$$
$\Phi,\Pi,c,M$ are regular as $\Lambda,L\to\infty$, 
$||M||<1$, and the property \r{7} is satisfied.

According to the previous section,
asymptotic solution to the 
Cauchy problem 
\r{54} at  $\tau=0$ is
$$
\matrix{
\Psi^{0}_g=U^g_{\Phi ',\Pi ',S'} c' \Lambda[p_1'(\cdot),q_1'(\cdot)]
...\Lambda[p_k'(\cdot),q_k'(\cdot)]\cr
\exp(\frac{1}{2}\int d{\bf k} a^+({\bf k})(M'a^+)({\bf k}))\Phi^{(0)},
}
\l{56}
$$
where the parameters 
entering to eq.
\r{56}, have the form
$$ \matrix{
 (p_j', q_j', \Phi ',\Pi ', R(M'), S', a(c'))=
{\cal U}^{0, -T_0}_{\xi(\cdot)}
{\cal U}^{-T_0, 0}_{0}
 (p_j, q_j, \Phi ,\Pi , R(M), 0, a(c)),\cr j=\overline{1,k},}
$$
$R(M)$ is the operator \r{39}, and $a(c)$ has the form
 \r{43}.

Since eqs. \r{25}, \r{26}, \r{33} are regular as
$\Lambda,L\to\infty$, parameters $\Phi', \Pi', S', p_j', q_j'$
are also finite in this limit. 

However, the integral \r{7} can be divergent in this limit, even if it was 
finite at

The operator $M'$ is a solution to eq. \r{6},
which is equal to
$M^{-T_0}$ at the initial time moment $t=-T_0$.
Remind that the integral \r{7} should be convergent at 
$t=-T_0$ even if one makes a limit $\Lambda,L\to\infty$.
However, the integral \r{7} can be divergent in this limit
at $t=0$.

Thus, we see that the Gaussian vector \r{23} of the regularized
theory
$$
u^g_{\Phi,\Pi,0} const \exp
(\frac{1}{2}\int d{\bf k} a^+({\bf k})(Ma^+)({\bf k}))\Phi^{(0)}
\l{57}
$$
correspond to renormalized state vector not if       
the quantity \r{7} is finite for the operator $M$, 
but if it is finite at $t<-T_0$ for the solution to eq.\r{35},
which is equal to $\tilde{M}$ at $t=0$.
The pre-factor of \r{57} is 
singular as $\Lambda,L \to \infty$; the dependence on
$\Lambda,L$ is determined from the following condition:
the solution to eq.
\r{36} which coincides with $c^0$ at  $t=0$ 
is regular at $t<-T_1$ as  $\Lambda,L \to \infty$.

Let us reformulate this statement for the operator $R$.
Consider the operator 
$\tilde{M}$ as a function of the coordinate and differential
operators:
$
\tilde{M}=
\tilde{M}(\qps{\bf x}{-i \partial/\partial{\bf x}}).
$
The function $\tilde{M}({\bf x},{\bf k})$ is called as a symbol
of the operator $\tilde{M}$. If the behaviour of the symbol 
at the infinity is the following
$$
\tilde{M}({\bf x}, {\bf k})=O(|{\bf k}|^{-d/2-\delta}).
$$
then the property \r{7}  is satisfied as 
$\delta>0$ and not satisfied as
$\delta \le 0$. 
It follows from the formula \r{39} that
the behavour of the symbol of the operator
$R$ at the infinity is              
$$
R({\bf x},{\bf k})=i\sqrt{{\bf k}^2+m^2} +
O(|{\bf k}|^{-d/2-\delta+1})
\l{58}
$$
At $\Lambda,L \to \infty$, the classical Hamiltonian system
\r{26} takes the form
$$
\ddot{\Phi}_c- \Delta \Phi_c + m^2 \Phi_c + \xi(t) V_{int}'(\Phi_c)=0
\l{59}
$$
while eq.\r{40} takes the form,
$$
\dot{R}+R^2+(-\Delta+m^2+\xi(t)V_{int}'' (\Phi_c(t,{\bf x}))=0
\l{60}
$$
Let   $\Phi(t,{\bf x})$ be a solution to eq.\r{59}
which satisfies the initial condition $\Phi_c |_{t=0}=\Phi$,
$\dot{\Phi}_c |_{t=0}=\Pi$. Consider the state vector of the   
regularized theory \r{57}. It corresponds to the following
functional $\Phi[\phi(\cdot)]$ in the Schr\"odinger representation,
$$ \matrix{
\Phi[\phi(\cdot)]= const \exp[\frac{i}{g}\Pi({\bf x})
(\phi({\bf x}) \sqrt{g} - \Phi({\bf x}))
\cr
+
\frac{i}{2} \int d{\bf x}
(\phi({\bf x})-\frac{\Phi({\bf x})}{\sqrt{g}})
(R(\phi-\frac{\Phi}{\sqrt{g}}))({\bf x}]
}
\l{61}
$$
The obtained result can be formulated as follows.

{\bf Statement 1.}
{\it
The Gaussian functional \r{61} corresponds to the renormalized state, 
if the solution to eq.\r{60},
which is equal to $R$ at $t=0$, satisfies at $t<-T_1$ the condition
\r{58} for $\delta>0$. If the property  \r{58} 
is satisfied for $\delta>0$ at $t<-T_1$, then the functional \r{61} 
does not correspond to any state vector of the renormalized
state space.
}

In the next sextion we will analyse eq.\r{60} and
formulate this condition in the more convenient form.

Let us consider the problem of regularity of the operator \r{48}
as $\Lambda,L\to \infty$.
Any vector of the Fock space can be approximated by the
linear combinations of vectors
$$
Y= c \Lambda[p_1(\cdot),q_1(\cdot)]
...\Lambda[p_k(\cdot),q_k(\cdot)]
\Phi^{(0)},
\l{62}
$$
Consider the vector \r{62} and analyse the 
problem of regularity of the vector
$$
\lim_{g\to 0} V^t_{\Lambda,L,g} Y
\l{63}
$$
at $\Lambda,L\to \infty$. Semiclassical  asymptotics of the vector
$$
 W^t_{\Lambda,L,g}
U^g_{\Phi^0,\Pi^0,0} Y =
(T_{\Lambda,L})^{-1} U^t_{\Lambda,L}
 T_{\Lambda,L}
U^g_{\Phi^0,\Pi^0,0} Y
$$
is constructed according to the previous section. It has the form
$
U^g_{\Phi',\Pi',S'}Y'
$, where
$$
Y'= c' \Lambda[p_1'(\cdot),q_1'(\cdot)]
...\Lambda[p_k'(\cdot),q_k'(\cdot)]
\exp(\frac{1}{2}\int d{\bf k} a^+({\bf k})(M'a^+)({\bf k}))\Phi^{(0)},
\l{64}
$$
The parameters entering to eq.\r{64} are
$$
\matrix{
 (p_j', q_j', \Phi ',\Pi ', R(M'), S', a(c'))=
\cr
{\cal U}^{0, -T_0}_{0}
{\cal U}^{-T_0, 0}_{\xi(\cdot)}
{\cal U}^{t, -T_0}_{\xi(\cdot)}
{\cal U}^{-T_0, 0}_{0}
 (p_j, q_j, \Phi^0 ,\Pi^0 , R(0), 0, a(c)), j=\overline{1,k},
}
\l{65}
$$
the function $\xi(t)$ is continued at $t>0$ as $\xi(t)=1$.
If 
$\Phi_{\lambda,L}^t=\Phi'$, $\Pi_{\Lambda,L}^t=\Pi'$,
$S_{\Lambda,L}^t=S'$, then the vector
 \r{63} coincides with the vector \r{64}.

Since  $p_j',q_j'$ are regular as
$\Lambda,L\to \infty$, 
it is sufficient to show that

(i) for $M'$ the integral \r{7} is finite;

(ii) $c'$ is regular as $\Lambda,L\to \infty$.

Note that the requirement (ii) is equivalent to the  
requirement that the 
quantity $ln a' = ln a(c')$ is finite if the condition (i)
is satisfied. The requirement  (i)
means that for the symbol 
of the operator $R'=R(M')$ the property 
\r{58} is satisfied.

Let us show that the conditions (i) and (ii) really take place,
if one chooses the counterterm  $H^{(1)}$ properly.

\centerline{{\bf 4. Singularities of the quadratic form}}

Consider  \r{60}. Denote by $R({\bf x}, {\bf k})$
the symbol of the operator $R$:
$$
R=R(\qp).
$$
Indices 1 and 2 mean that the differential operators act first and 
multiplication operators act next:
$$
R \int d{\bf k} f({\bf k}) e^{i{\bf kx}}
\equiv \int d{\bf k} R({\bf x},{\bf k}) f({\bf k})
e^{i{\bf kx}}.
\l{66}
$$
In this section we investigate the asymptotics of the
symbol $R({\bf x},{\bf k})$ at ${\bf k} \to \infty$.

We consider the asymptotic expansion of the sumbol in    
$1/\omega_{\bf k}$, $\omega_{\bf k} = \sqrt{{\bf k}^2+m^2}$
at  ${\bf k} \to \infty$:
$$
R^t({\bf x},{\bf k})=
{\cal R}^t({\bf x},{\bf k})
=i\omega_{\bf k} + \sum_{m\ge 1}
\frac{R^t_m({\bf x},{\bf k}/\omega_{\bf k})}{\omega_{\bf k}^m},
\l{67}
$$
where $R_m^t({\bf x},{\bf n})$ are smooth functions, rapidly damping
at ${\bf x}\to \infty$ with all their derivatives.
For the case of ${\bf x}$- independent fields, such an expansion
was considered in ref.\c{9}.

Let us substitute eq.\r{67} to eq.\r{60} and find the conditions
which are necessary to make the left-hand side of eq.\r{60} 
of order
$O(\frac{1}{\omega_{\bf k}^j})$, where $j$ -- is an arbitrary
number.

Let us use the formula for the symbol of the product of the   
operators (see, for example, \c{28,29}). Let
$\hat{A}=A(\qp)$, $\hat{B}=B(\qp)$ be operators.
Then their product    $\hat{A}\hat{B}$ has the following symbol
 $A*B$,
$$
\hat{A}\hat{B}=A*B(\qp),
$$
which is presented as follows:
$$
(A*B)({\bf x},{\bf k})= A(\qps{\bf x}{{\bf k}
-i\frac{\partial}{\partial {\bf x}}) B({\bf x},{\bf k}}).
\l{68}
$$
$$
(A*B)({\bf x},{\bf k})=\int \frac{d{\bf z} d{\bf p}}{(2\pi)^d}
A({\bf x},{\bf k}-{\bf p}) B({\bf x}+{\bf z},{\bf k})
e^{i{\bf pz}}
\l{69}
$$
To obtain eq.\r{69}, it is sufficient to make use of the relation
$$
\hat{B} \int d{\bf p} f({\bf p})e^{i{\bf px}} =
\int d{\bf k} e^{i{\bf kx}}
\int \frac{d{\bf z} d{\bf p}}{(2\pi)^d}
e^{i({\bf p}-{\bf k}) {\bf y}} B({\bf y},{\bf p})
f({\bf p}),
$$
and of the formula \r{66}. 
Formula \r{68} is a corollary of eq.\r{69}.

Perform the formal expansion of the right-hand side 
of the formula \r{68}. One obtains:
$$
(A*B)({\bf x},{\bf k})= \sum_{l\ge 0}
\frac{(-i)^l}{l!}
\frac{\partial^l A({\bf x},{\bf k})}
{\partial k_{i_1}...\partial k_{i_l}}
\frac{\partial^l B({\bf x},{\bf k})}
{\partial x_{i_1}...\partial x_{i_l}},
\l{70}
$$
where we sum over repeated indices.
If $A$ and $B$ are
$$
A=
\frac{A_1^t({\bf x},{\bf k}/\omega_{\bf k})}{\omega_{\bf k}^{m_1}},
B=
\frac{B_{1}^t({\bf x},{\bf k}/\omega_{\bf k})}{\omega_{\bf k}^{m_2}},
$$
the series \r{70} is an asymptotic series in terms of    
$\frac{1}{\omega_{\bf k}}$, because the $l$-th term of
formula \r{70} is of order $O(\omega_{\bf k}^{-m_1-m_2-l})$
at ${\bf k}\to \infty$. One can prove eq.
\r{70} 
by applying the stationary phase method
 \c{30} to the integral \r{69}.

Making use of formula \r{70}, substituting the symbol $R*R$ 
of the operator
$R^2$ to eq.\r{60} and considering the terms of order
 $O(1/\omega_{\bf k}^s)$, $s\ge 0$,
we find:
$$
2i R_1^t({\bf x},{\bf n})+ \xi(t) V_{int}''(\Phi^t({\bf x}))=0
$$
at $s=0$ and
$$
\dot{R}^t_s({\bf x},{\bf n})+ 2iR^t_{s+1}({\bf x},{\bf n})
+\sum_{{l+m=s+1}\atop {l,m\ge 1}} \omega_{\bf k}^{l-1}
\frac{(-i)^{l-1}}{l!}
\frac{\partial^l \omega_{\bf k}}
{\partial k_{i_1}...\partial k_{i_l}}
\frac{\partial^l R^t_m({\bf x},{\bf n})}
{\partial x_{i_1}...\partial x_{i_l}},
$$
$$
+ \sum_{{m+n+l=s } \atop {m,n \ge 1,l\ge 0} }
\frac{(-i)^l}{l!} \omega_{\bf k}^{m+l}
\frac{\partial^l}
{\partial k_{i_1}...\partial k_{i_l}}
\frac{R^t_m({\bf x},{\bf k}/\omega_{\bf k})}{ \omega_{\bf k}^{m} }
\frac{\partial^l R_n^t({\bf x},{\bf n})}
{\partial x_{i_1}...\partial x_{i_l}}=0,
\l{71}
$$
at  $s\ge 1$;
in this formula ${\bf n}={\bf k}/\omega_{\bf k}$.
We have used the assumption that
quantities $\dot{R}_s^t$ and $R_s^t$ are of the same order.

Recursive relation \r{71} allows us to express
$R_{s+1}^t$ through the previous orders of the perturbation 
expansion in $1/\omega_{\bf k}$, $R_1^t,...,R_s^t$. 
Note that the constructed function
 $R_{s+1}^t$ is determined uniquely if one knows the values
 of the field
$\Phi^t(\cdot)$ and momenta $\Pi^t(\cdot)$ at this time moment
$t$ and values of the function $\xi$  and its derivatives
at time $t$.

If $\xi(t)=0$, â.¥. $t<-T_1$, the obtained coefficients
are also equal to zero, so that
$R^t=i\omega_{\bf k}$ at these values of $t$. 
If $t>-T_2$ the asymptotics is $\xi$-independent.

The first functions $R_s^t$ defined from
eqs. \r{71} have the form:
$$
R^t_1({\bf x},{\bf n})= \frac{i}{2}\xi(t) V_{int}''(\Phi^t({\bf x})),
$$
$$
R_2^t({\bf x},{\bf n})=-\frac{1}{4}
\left(\frac{\partial}{\partial t}+{\bf n}
\frac{\partial}{\partial {\bf x}}\right)
\xi(t) V_{int}''(\Phi^t({\bf x})).
$$
Note that the first derivative of  $\Phi^t$ 
is expressed through the momentum 
$\Pi^t$, while next derivatives entering to next functions
$R_s$, can be expressed from classical equations of motion
through      $\Phi^t$ and $\Pi^t$.

We have constructed the asymptotics at $|{\bf k}| \to \infty$ only
for one solution to eq.\r{60} corresponding to
vacuum at $t < -T_1$, because it is equal at these values of 
$t$ to $i\omega_k$. Consider the construction 
of the asymptotic solution to eq.\r{60} which obeyes the initial
condition \r{58}. Consider the following substitution to
eq.\r{60}:
$$
R^t={\cal R}^t+r^t,
$$
where ${\cal R}^t$ is the constructed solution
\r{67}.
Considering the leading order 
in $1/\omega_{\bf k}$, one obtains that:
$$
\dot{r}^t+ 2i\omega_{\bf k} r^t=0.
$$
The solution to this equation is 
$r^T=r^{-T_0} \exp(-2i\omega_{\bf k}(t+T_0))$;
the corrections are constructed in \c{31}.

Thus, statement 1 can be reformulated as follows.

{\bf Statement  2.} {\it
The Gaussian functional \r{61} correspond to the
renormalized state if and only if
 $R^0$ is equal to ${\cal R}^0$ (eq.\r{67}) up to
$O(|k|^{-d/2+1-\delta})$ 
$$
  R^0={\cal R}^0+O(|k|^{-d/2+1-\delta}),
\l{72}
$$
The requirement \r{72} does not depend on choice 
of the function 
$\xi(t)$ and depends on the functions    $\Pi^0({\bf x})$ and
$\Phi^0({\bf x})$ only.
}

Let us check the conditions (i) and (ii) formulated
for $R'$ and $a'$ at the end of the previous section.

The condition on  $R'$ is a corollary of the property
of invariance of the asymptotics
\r{67} under substituting the function $\xi(t)$  by the function
$\xi(t-t_0)$, where $t_0=const$.

The quantity $a'$ is finite
if the counterterm 
$H^{(1)}$ compencate the divergences arising 
when one calculates the trace of $R$.
To find the form of the counterterm, one should
consider asymptotic solutions of the regularized 
equation 
\r{40}:
$$
{\cal R}^t_{\Lambda,L} ({\bf x},{\bf k})=
i\omega_{\bf k} + \sum_{m\ge 1, j \ge 0}
\frac{R^t_{mj}({\bf x},{\bf k}/\omega_{\bf k}
,{\bf k}/\Lambda)}{\omega_{\bf k}^m \Lambda^j},
$$
extract the divergence in the trace 
$$
Tr ( Im {\cal R}^t_{\Lambda,L} - i\omega_k)
\sim \sum_{m=\overline{1,d},j=\overline{0,d}}
\int
\frac{ d{\bf k} d{\bf x} Im R^t_{mj}({\bf x},{\bf k}/\omega_{\bf k}
,{\bf k}/\Lambda)}{\omega_{\bf k}^m \Lambda^j},
$$
add it to the divergence part of the trace
$\frac{1}{2}Tr \frac{1}{\hat{\omega}} 
\frac{\partial^2H^{(0)}}{\partial \Phi \partial \Phi}
$
and consider  $H^{(1)}$ to be equal to this sum. 
Since $R^t_{mj}$
depends only on $\Phi^t(\cdot)$, $\Pi^t(\cdot)$
and on values of the function  $\xi$
and its derivatives at time moment  $t$, 
the counterm $H^{(1)}$ is a functional of the
field and of the momenta at $t>-T_2$. 
Thus, we have checked the requirements (i) and (ii).

\centerline{{\bf 5. Conclusion}}

Thus, we have found the condition formulated in
statements 1 and 2 instead of eq.\r{7}.
If this condition is satisfied, the Gaussian functional
\r{61} corresponds to an element of the renormalized state 
space. 

Let us compare the obtained condition on the quadratic form entering
to the Gaussian vector with the prescriptions formulated in other
papers on the semiclassical field theory. It is usually non-manifestly
assumed in the papers that do not contain the analysis of renormalization
that the renormalized states coincide with states in the regularized
field theory, so that one should use the condition \r{7}. This means
that the requirement \r{58} should be imposed on $R$.
However, this condition may be non-invariant under time evolution.

It was suggested in ref.\c{32} 
that one should consider initial states which are eigenstates 
for the Hamiltonian operator at the initial time moment. 
Such states correspond to the Gaussian functionals with the following
singular part of the operator
 $R^t$:
$$
(R*R)({\bf x},{\bf k}) \sim -({\bf k}^2 + m^2 + V''(\Phi_c^t({\bf x})))
\l{73}
$$
Formulas \r{58} and \r{73} are colloraries of our condition
\r{72} if and only if the asymptotics is constucted 
as
$|{\bf k}|\to
\infty$ up to                $O(1/|{\bf k}|)$
and $O(1/|{\bf k}|^2)$ accuracy correspondingly.
For general non-stationary solutions, this is correct
only for sufficiently small number of the space-time dimensions.
On the other hand, for stationary solutions the condition
\r{73} is always correct.

Although we have considered the case of the single scalar field
only, the more complicated cases can be studied analogously.

For gauge theories \c{33}, one can consider the Coulomb gauge
and write the evolution equation in this gauge. Then the 
Bogoliubov's procedure of ''switching on'' the interaction
can be applied.

For theories with fermions, one can also consider the ansatz 
\r{23}, where the operator $U$ contains the boson fields only.
Eq.\r{23} will obey the quantum equation of motion
if the state vector $Y^t$ satisfies the fermionic Schr\"odinger
equation with the quadratic Hamiltonian. This is an exactly
solvable equation \c{12}.

The obtained condition on the operator $R$ does not depend 
on counterterms; it depends on the classical action only.
Thus, the obtained requirement on $R$ can be written both
for renormalizable and non-renormalizable cases.

The pre-exponential factor depends on the one-loop counterterm
 $H^{(1)}$. This is in agreement with the well-known calculations
 of the functional integral in quantum field theory by the 
 saddle-point technique: tree Feynman graphs correspond to 
 the classical action, while one-loop graphs correspond to the
 determinat of fluctuations.

 The problem of divergences and renormalization in quantum field
 theory is usually associated with loop Feynman graphs. We see from
 the obtained results that renormalization is very important 
 even at the tree level, i.e. in the classical field theory.
 It is the problem of divergences and renormalization that
 leads us to the conditions on the solution to eq.\r{60}. This
equation is classical, not quantum, because it does not depend on
 the counterterms.

We have also seen that the pre-exponential factor in eq.\r{57}
can be divergent, because the counterterm 
$H^{(1)}$ compensates the infinite phase of $c^t$, while
the divergence in $|c^t|$ which arises from the combination
of determinants \r{43} cannot be removed.
This observation can be interpretted as follows.
It is known that the Heisenberg fields $\hat{\varphi}({\bf x},t)$
are well-defined operatoer distributions 
(i.e. expression $\int dt d{\bf x} \hat{\varphi}({\bf x},t)
f({\bf x},t)$, 
where $f$ is a c-number function,
determines the operator). However,
Schr\"odinger field operators may not be defined as operator 
distributions 
 \c{34,13,35}
(i.e. expression $\int  d{\bf x} \hat{\varphi}({\bf x},t)
f({\bf x})$  is not a well-defined operator).
This means that the interpretation of the Schr\"odinger 
functional as a probability amplitude that the value of field 
$\varphi$ is given is not consistent. On the other hand, the notion of
the quadratic form entering to the wave functional can be 
introduced.

We have considered the leading order of the semiclassical
approximation only. One can investigate larger orders in
analogous way. However, one should take into account
the volume divergences corresponding to the Haag theorem
 \c{35,34}. This means that the transformation $T_{\Lambda,L}$
 should be choosen as a composition of the constructed
 transformation \r{50} and the Faddeev transformation
 \c{14} which removes the vacuum divergences.

The problem of regularity of the renormalized evolution
operator at larger orders of the semiclassical expansion can be
reduced to the problem of regularity of the 
Bogoliubov's S-matrix corresponding to the 
interaction which is "swithched" off at the infinity.
If the classical solution vanishes, the problem can be
solved by using the Bogoliubov-Parasiuk theorem
\c{16,36,37}.

We will consider these problems in details in our following
publications.

\centerline{{\bf REFERENCES}}

\i{1}
R. Dashen, B. Hasslasher and A. Neveu 
 {\it Phys. Rev.}{\bf  D10} (1974) 4114
\i{2}
L.D.Faddeev and V.E.Korepin {\it Phys. Rep.}{\bf 42} (1978) 1
\i{3} 
R.Rajaraman,
{\it Solitons and Instantons. An Introduction to solitons and instantons in
quantum field theory.} (Amsterdam, Netherlands: North-Holland  1982).
\i{4} 
 A.A. Grib, S.G. Mamaev, and V.M. Mostepanenko,
{\it Vacuum Quantum Effects in Strong  Fields} (Atomizdat, Moscow,
1988; Friedmann Laboratory Publishing, St. Petersburg 1994).
\i{5} 
N.D. Birrell, P.C.W. Davies, 
{\it Quantum Fields in Curved Space.}
(Cambridge, UK: Univ. Pr. , 1982).
\i{6}
E.S. Fradkin, D.M. Gitman and Sh.M. Shvartsman,
{\it  Quantum Electrodynamics with Unstable Vacuum.}
(Berlin, Germany: Springer ,1991).
\i{7}
V.G.Bagrov, D.M.Gitman and V.A.Kuchin,
In: {\it Urgent problems of theoretical physics},
(ed. A.A.Sokolov, Moscow Univ. Press, 1976.)
\i{8} F. Cooper and E. Mottola
{\it Phys.Rev.} {\bf D36} (1987) 3114
\i{9}
S.-Y.Pi and M.Samiullah {\it Phys. Rev.}{\bf  D36} (1987) 3128
\i{10}
O. Eboli, S.-Y. Pi and M. Samiullah
{\it Ann. Phys.} {\bf 193} (1989) 102
\i{11}
A.A.Grib and S.G.Mamaev, {\it Yadernaya Fizika} {\bf 10} (1969) 1276
\i{12}
F.A. Berezin. {\it The method of second quantization.}
(Moscow: Nauka, 1965; N.Y., 1966).
\i{13}
O.I. Zavialov and V.N.Sushko,
In: {\it Statistical Physics and Quantum Field Theory},
ed. N.N.Bogoliubov (Moscow: Nauka, 1973).
\i{14}
L.D. Faddeev, {\it Doklady Akad. Nauk SSSR} {\bf 152}
(1963) 573.
\i{15} N.N. Bogoliubov, {\it Doklady Akademii Nauk SSSR}
{\bf 81} (1951) 757.
\i{16}
N.N. Bogoliubov and D.V. Shirkov, {\it Introduction to the
Theory of Quantized Fields} (N.-Y.: Interscience Publishers, 1959).
\i{17} V.P.Maslov. {\it The complex-WKB method for
nonlinear equations} (Moscow: Nauka, 1977).
\i{18}
F. Cooper, S.-Y. Pi and P.Stancioff, {\it Phys. Rev.}
{\bf D34} (1986) 3831.
\i{19}
V.P.Maslov and O.Yu.Shvedov, {\it Teor. Mat. Fiz.} {\bf 104} (1995) 310
\i{20}
V.P.Maslov and O.Yu.Shvedov, {\it Teor. Mat. Fiz.} {\bf 104} (1995) 479
\i{21}
E.C.G. Stueckelberg  {\it Phys.Rev.} {\bf 81} (1951) 130.
\i{22} K. Hepp, {\it Theorie de la renormalisation}
(Springer-Verlag, 1969).
\i{23}
J. Glimm and A. Jaffe,
{\it Boson quantum field models.}
In ''London 1971, Mathematics Of Contemporary Physics'', London 1972,
pp. 77-143.
\i{24}
I. Ya. Arefieva, {\it Teor. Mat.Fiz.} {\bf 14} (1973) 3
\i{25}
I. Ya. Arefieva, {\it Teor. Mat.Fiz.} {\bf 15} (1973) 207
\i{26}
A.S. Shvarts, {\it Mathematical Foundations of
Quantum Field Theory} (Moscow: Atomizdat, 1975).
\i{27}
A.D. Sukhanov, {Zh.Eksp.Teor.Fiz.} {\bf 43} (1962) 932.
\i{28}
M.V.Karasev and V.P.Maslov, {\it Nonlinear Poisson
Brackets. Geometry and Quantization} (Moscow: Nauka, 1991).
\i{29}
V.P.Maslov. {\it Operational Methods}. (Moscow: Mir Publishers, 1976).
\i{30}
V.P.Maslov. {\it Perturbation Theory and Asymptotic Methods}.
(Moscow: Moscow Univ. Press, 1965).
\i{31} V.P. Maslov and O.Yu. Shvedov,
{\it The complex-WKB method for relativistic field theory},
submitted to Doklady Akademii Nauk.
\i{32}
M.I. Shirokov, {\it Yadernaya Fizika} {\bf 7} (1968) 672.
\i{33}
A.A. Slavnov and L.D.Faddeev,
{\it Introduction to the Quantum Theory of Gauge Fields}
(Moscow: Nauka, 1978).
\i{34}
A. Wightmann {\it Problems in Relativistic Theory of
Quantized Fields} (Moscow: Mir, 1968).
\i{35}
N.N. Bogoliubov, A.A. Logunov, A.I.Oksak and
I.T.Todorov.
{\it General Principles of Quantum Field Theory}.
(Moscow: Nauka, 1987).
\i{36} O.I. Zavialov.
{\it Renormalized Feynman Graphs.} (Moscow: Nauka, 1979).
\i{37}
A.S. Shvarts, {\it Elements of Quantum Field Theory.
Boson Interactions} (Moscow: Atomizdat, 1975).

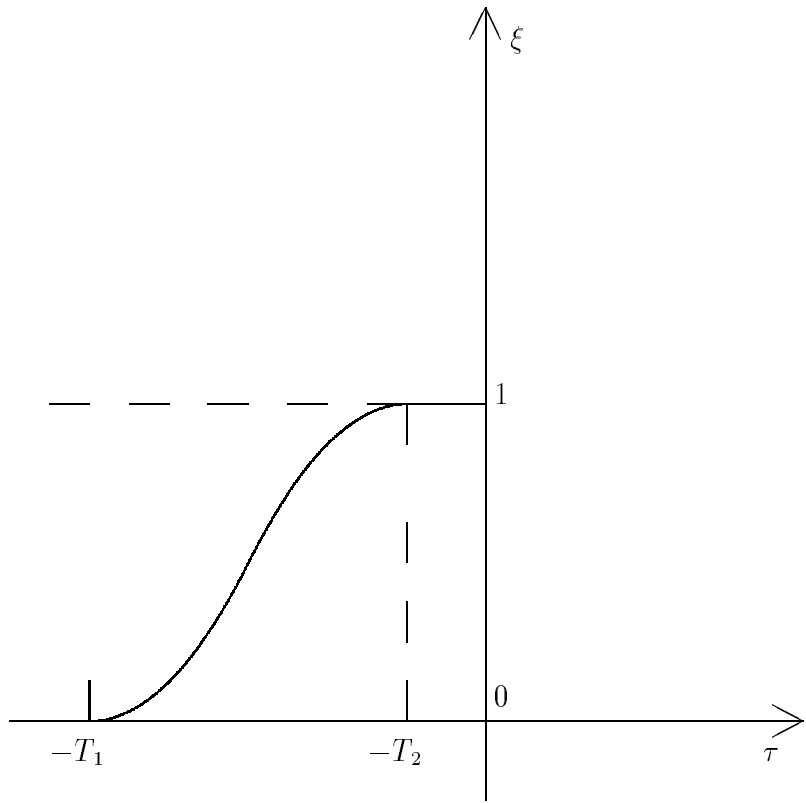
\end